\documentclass{ckm}                 

\usepackage{mathptmx}          

\RequirePackage{xspace}
\usepackage{relsize}
\def\babar{\mbox{\slshape B\kern-0.1em{\smaller A}\kern-0.1em
    B\kern-0.1em{\smaller A\kern-0.2em R}}}
\def\rar{\rightarrow}
\def\Bz      {\ensuremath{B^0}\xspace}
\def\Bzb     {\ensuremath{\Bbar^0}\xspace}
\def\BzBzb   {\ensuremath{\Bz {\kern -0.16em \Bzb}}\xspace}
\def\Abar{\kern 0.18em\overline{\kern -0.18em A}{}}
\def\Apm{A^{+-}}
\def\Amp{A^{-+}}
\def\Apmb{\Abar^{+-}}
\def\Ampb{\Abar^{-+}}

\def\Apz{A^{+0}}

\def\Apzb{\Abar^{+0}}
\def\Azp{A^{0+}}
\def\Azpb{\Abar^{0+}}
\def\Azz{A^{00}}
\def\Azzb{\Abar^{00}}
\def\Tpm{T^{+-}}
\def\Tmp{T^{-+}}
\def\Ppm{P^{+-}}
\def\Pmp{P^{-+}}
\def\Tzz{T^{00}}
\def\Tpz{T^{+0}}

\def\lampm{\lambda^{+-}}
\def\lammp{\lambda^{-+}}
 \def\lampmmp{\lambda^{+-(-+)}}
\def\kappm{\kappa^{+-}}
\def\kapmp{\kappa^{-+}}
\def\kappmmp{\kappa^{+-(-+)}}
\def\lamCP{\lambda_{\rm CP}}
\def\lamFL{\lambda_{\rm tag}}

\def\alphaeffpmmp{\alpha_{\rm eff}^{+-(-+)}}

\def\Bbar    {\kern 0.18em\overline{\kern -0.18em B}{}\xspace}

\def\BRpmb{{\cal \kern 0.18em\overline{\kern -0.18em  B}}{}_{\rho\pi}^{+-}}
\def\BRmpb{{\cal \kern 0.18em\overline{\kern -0.18em  B}}{}_{\rho\pi}^{-+}}

\def\BRipmb{{\cal \kern 0.18em\overline{\kern -0.18em  B}}{}_{\rho^+\pi^-}}
\def\BRimpb{{\cal \kern 0.18em\overline{\kern -0.18em  B}}{}_{\rho^-\pi^+}}
\def\BRall{{\cal B}_{\rho\pi}^{\pm\mp}}

\def\BRpz{{\cal B}_{\rho\pi}^{+0}}
\def\BRzp{{\cal B}_{\rho\pi}^{0+}}
\def\BRzz{{\cal B}_{\rho\pi}^{00}}
\def\BRrhoK{{\cal B}_{\rho K}^{\pm\mp}}

\def\Czz{C_{\rho\pi}^{00}}
\def\Szz{S_{\rho\pi}^{00}}
\def\Crhopi{C_{\rho\pi}}
\def\dCrhopi{\Delta C_{\rho\pi}}

\def\Srhopi{S_{\rho\pi}}
\def\dSrhopi{\Delta S_{\rho\pi}}
\def\CrhoK{C_{\rho K}}
\def\dCrhoK{\Delta C_{\rho K}}
\def\SrhoK{S_{\rho K}}
\def\dSrhoK{\Delta S_{\rho K}}

\confname{Workshop on the CKM Unitarity Triangle, IPPP Durham, April
  2003}

\title{$B^0 \rightarrow \pi^+ \pi^- \pi^0$ feasibility studies}

\author{J. Stark}
\address{Laboratoire de Physique Subatomique et de Cosmologie, Grenoble, France}

\begin{document}

\begin{abstract}
The potential of constraints on the angle~$\alpha$ of the Unitarity Triangle from studies of $B^0/\overline{B}^0 \rightarrow \pi^+ \pi^- \pi^0$ decays is reviewed. The experimental inputs needed for an isospin analysis are starting to become available. The precision of current measurements is extrapolated to higher luminosities and the constraints on~$\alpha$ that could be obtained in the foreseeable future are discussed. Studies of the sensitivity of a time-dependent Dalitz plot analysis are briefly reviewed.
\end{abstract}

\maketitle

\section{Introduction}
The \babar\ and Belle Collaborations have performed searches for CP~violation in $B^0/\overline{B}^0$~decays to~$\pi^+ \pi^-$~\cite{bib:pipi}, where the mixing-induced CP~asymmetry is related to the angle~$\alpha$ of the Unitarity Triangle~(UT). The implications of these results are discussed in the contribution~\cite{bib:Lydia} to the present proceedings. In this paper, we discuss the $\pi^+ \pi^- \pi^0$~final state which probes both direct and mixing-induced CP~violation. The mixing-induced CP~violation is related to~$\alpha$.

\begin{figure}
\hbox to\hsize{\hss
\includegraphics[width=0.4\hsize]{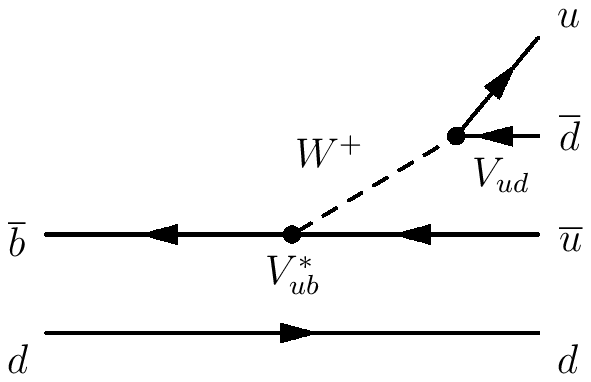}
\hss
\includegraphics[width=0.4\hsize]{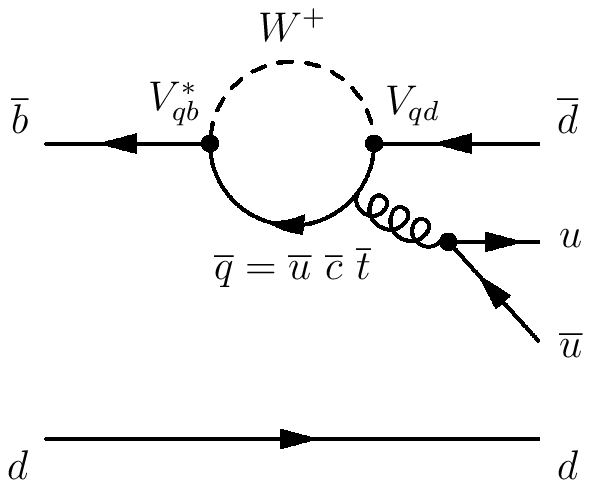}
}
\vspace{3mm}
\hbox to\hsize{\hss
\includegraphics[width=0.4\hsize]{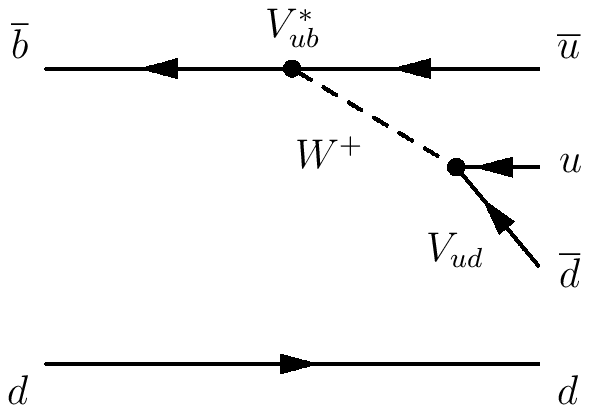}
\hss
}
\vskip-5cm
~~~~~(a)~~~~~~~~~~~~~~~~~~~~~~~~~~~~~~~~~~~~~~~~(b)
\vskip2.5cm
~~~~~~~~~~~~~~~~~(c)
\vskip1.6cm
\caption{Feynman diagrams for $\Bz \rightarrow \rho^{\pm} \pi^{\mp}$ and $\Bz \rightarrow \rho^0 \pi^0$.}
\label{fig:Feynman}
\end{figure}

The decay $B^0/\overline{B}^0 \rightarrow \pi^+ \pi^- \pi^0$ is dominated by the resonant intermediate states~$\rho^+ \pi^-$ and~$\rho^- \pi^+$~\cite{bib:rhopibfCLEO,bib:rhopibfBabar,bib:rhopibfBelle}. The leading Feynman diagrams for the $B^0 \rightarrow \rho^{\pm} \pi^{\mp}$ decays are shown in Fig.~\ref{fig:Feynman}(a). Note that Fig.~\ref{fig:Feynman}(a) represents two distinct diagrams, one where the $\rho$~meson is formed from the $W^+$~boson, and one where the $\rho$~meson is formed with the spectator quark. As pointed out in~\cite{bib:Royetal}, these diagrams mean that both $B^0$~and~$\overline{B}^0$ can decay to the $\rho^+ \pi^-$~state, and therefore that CP~violation can manifest itself in the interference between decays with and without mixing. The moduli of the two corresponding amplitudes are expected to be comparable in size, and CP-violating effects could thus be large in these decays. The same comment applies to the $\rho^- \pi^+$~state. Contributions from penguin diagrams (see Fig.~\ref{fig:Feynman}(b)) could be sizeable and complicate the interpretation of measured CP~asymmetries in terms of the elements of the CKM~matrix. The decay $B^0/\overline{B}^0 \rightarrow \pi^+ \pi^- \pi^0$ can also proceed through the $\rho^0 \pi^0$ intermediate state. The corresponding (colour suppressed) Feynman diagram is shown in Fig.~\ref{fig:Feynman}(c). This decay has not been observed yet; current experimental limits are of the order of~$5 \cdot 10^{-6}$~\cite{bib:rhopibfCLEO,bib:rhopibfBabar,bib:rhopibfBelle}. Contributions from scalar resonances like, e.g., $f_0(980) \rightarrow \pi^+ \pi^-$ and higher excitations are also expected. Non-resonant contributions to $B^0/\overline{B}^0 \rightarrow \pi^+ \pi^- \pi^0$ are small, they are found to represent less than $4 \; \%$ of the total rate~\cite{bib:rhopibfBelle}.

Different approaches to the analysis of $B^0/\overline{B}^0 \rightarrow \pi^+ \pi^- \pi^0$ are used in the literature. In the so-called quasi-two-body approach~\cite{bib:PhysBook} one restricts oneself to the two regions of the $\pi^+ \pi^- \pi^0$~Dalitz plot dominated by either $\rho^+ \pi^-$ or $\rho^- \pi^+$. Another approach is the time-dependent Dalitz plot analysis originally proposed in~\cite{bib:DalitzAnal}. It is optimal in the sense that it exploits all available information, but experimentally much more involved. Specifically, the full dynamics of the $\pi^+ \pi^- \pi^0$ final state are studied, and the interference between the $\rho^+ \pi^-$, $\rho^- \pi^+$ and $\rho^0 \pi^0$ contributions in different points of the Dalitz plot is used to obtain simultaneous constraints on the weak and strong phases in these transitions. This analysis allows the measurement of~$\alpha$ even in the presence of non-negligible penguin contributions.

The quasi-two-body approach is discussed in section~\ref{sec:twobody}. Recent \babar\ results obtained in that framework are reviewed. Based on this and other experimental results, the sensitivity to~$\alpha$ is extrapolated to higher luminosities expected to be available in the future. Present data samples are not large enough for a time-dependent Dalitz plot analysis. The expected statistical sensitivity and some of the major difficulties expected in this analysis are briefly discussed in section~\ref{sec:Dalitz}.

\section{Quasi-two-body analysis}
\label{sec:twobody}
The observables in this analysis are defined in section~\ref{sec:definitions}. A recent \babar\ measurement of these observables is summarised in section~\ref{sec:measurement}. In section~\ref{sec:constraints}, the present sensitivity to~$\alpha$ is assessed and extrapolated to higher luminosities. The discussion in sections \ref{sec:definitions}~and~\ref{sec:constraints} follows essentially the discussion in~\cite{bib:LALnote}, and the reader is referred to this paper for additional details.

\subsection{Basic formulae and definitions}
\label{sec:definitions}
The four decay amplitudes relevant for the quasi-two-body analysis can be written as
\begin{equation}
\label{eq:b0rho+pi-}
\begin{array}{l}
\Apm \equiv  A(\Bz\rar\rho^+\pi^-)
                        =   R_u e^{+i\gamma}\Tpm
                            + R_t e^{-i\beta}\Ppm~, \\
\Amp \equiv  A(\Bz\rar\rho^-\pi^+)
                        =   R_u e^{+i\gamma}\Tmp
                            + R_t e^{-i\beta}\Pmp~, \\
\Apmb \equiv  A(\Bzb\rar\rho^+\pi^-)
                        =   R_u e^{-i\gamma}\Tmp
                            + R_t e^{+i\beta}\Pmp~, \\
\Ampb \equiv  A(\Bzb\rar\rho^-\pi^+)
                        =   R_u e^{-i\gamma}\Tpm
                            + R_t e^{+i\beta}\Ppm~, \\
\end{array}
\end{equation}
where the $T$~and~$P$ symbols represent the effective complex tree and penguin contributions. Contributions with other weak phases are neglected here. The parameters $R_u = |V_{ud}V^*_{ub}|$, $R_c = |V_{cd}V^*_{cb}|$ and $R_t = |V_{td}V^*_{tb}|$ are the sides of the~UT, $\beta$ and $\gamma$ follow the usual convention for the angles of the~UT. The unitarity relation $R_u e^{i\gamma}+R_c e^{i\pi}+R_t e^{i\beta}=0$, valid to~${\cal O}(\lambda^5)$ in the Wolfenstein expansion parameter~$\lambda$, has been used to express the contributions from $c$~quark loops in the penguin amplitudes in terms of the other contributions.

We define
\begin{equation}
\label{eq:lambdakappa}
\begin{array}{c}
        \lampm \equiv \frac{q}{p}\frac{\Apmb}{\Apm}~,\hspace{1.cm}
        \lammp \equiv \frac{q}{p}\frac{\Ampb}{\Amp}~,\\
        \kappm \equiv \frac{q}{p}\frac{\Ampb}{\Apm}~,\hspace{1.cm}
        \kapmp \equiv \frac{q}{p}\frac{\Apmb}{\Amp}~,
\end{array}
\end{equation}
where the parameters $p$~and~$q$ describe the composition of the $B^0$~mass eigenstates in terms of the flavour eigenstates~\cite{bib:PhysBook}. These definitions are inspired by the definition of the parameter~$\lambda$ that is commonly used in the discussion of $B$~decays to CP~eigenstates (see, e.g.~\cite{bib:PhysBook}). Each of the two $\lampmmp$~parameters defined above involves only one $\rho \pi$~charge combination. They are insensitive to direct CP~violation, but their imaginary part is the sum of contributions from the weak phase~$\alpha$ and strong phases. The $\kappmmp$~parameters involve both $\rho \pi$~charge combinations, but only one of the two amplitudes represented in Fig.~\ref{fig:Feynman}(a), i.e. either corresponding to transitions where the~$\rho$ is formed from the~$W$ or from the spectator quark. Their moduli are linked to direct CP~violation, while their phases measure effective weak angles~$\alphaeffpmmp$ defined below. The parameters~$\lampmmp$ do not have the same properties as the parameter~$\lambda$ in the analysis of CP~eigenstates, i.e. $\lampm \neq \pm 1$ or~$\lammp \neq \pm 1$ does not necessarily imply CP~violation. We define the parameters~$\lamCP$ and~$\lamFL$ such that
\begin{eqnarray*}
\label{eq:dt}
        |\lamCP|^2
        &\equiv&
        \frac{|\lampm|^2 + |\lammp|^2 + 2 |\lampm|^2 |\lammp|^2}
             {2 + |\lampm|^2 + |\lammp|^2}~,\\[0.3cm]
        |\lamFL|^2
        &\equiv&
        \frac{1 + 2|\lampm|^2 + |\lampm|^2 |\lammp|^2}
             {1 + 2|\lammp|^2 + |\lammp|^2 |\lampm|^2}~,\\[0.3cm]
        {\rm Im}\lamCP
        &\equiv&
        \frac{{\rm Im}\lampm(1 + |\lammp|^2) + {\rm Im}\lammp(1 + |\lampm|^2)}
             {2 + |\lampm|^2 + |\lammp|^2}~,\\[0.3cm]
        {\rm Im}\lamFL
        &\equiv&
        \frac{{\rm Im}\lampm(1 + |\lammp|^2) - {\rm Im}\lammp(1 + |\lampm|^2)}
             {1 + 2|\lammp|^2 + |\lammp|^2 |\lampm|^2}~.
\end{eqnarray*}
$\lamCP$~has the desired properties, i.e.~$\lamCP \neq 1$ implies CP~violation. $\lamFL$~is a measure of the ``{\it self-taggedness}'' of the $\Bz/\Bzb \rightarrow \rho^{\pm} \pi^{\mp}$ decay. If, e.g., $\Apm = \Ampb = 0$, i.e. the charge of the~$\rho$ identifies the flavour of the~$B$ (the $B \rightarrow \rho \pi$ decay is {\it self-tagging}), then~$\lamFL = 0$ and mixing-induced CP~violation cannot occur. For the other extreme case, $|\Apm| = |\Amp|$ and $|\Apmb| = |\Ampb|$, we obtain~$|\lamFL|=1$.

At an asymmetric $B$-factory like PEP-II/\babar, the time-dependence of $\Bz/\Bzb \rightarrow \rho^{\pm} \pi^{\mp}$ decays is studied using pairs of $B$~mesons from $\Upsilon(4S) \rightarrow \BzBzb$, where one of the $B$~mesons (denoted $B_{\rho \pi}$) is reconstructed in the $\rho \pi$~final state, and the flavour of the other $B$~meson (denoted $B_{\rm tag}$) is determined ({\it tagged}) from its decay products. Defining $\Delta t = t_{\rho \pi} - t_{\rm tag}$ as the proper time interval between the decays of the~$B_{\rho \pi}$ and the~$B_{\rm tag}$, the time-dependent decay rates are given by
\begin{eqnarray}
\label{eq:thTime}
  \lefteqn{f^{\rho^\pm \pi^\mp}_{Q_{\rm tag}}(\Delta t) = (1\pm A_{\rm CP}^{\rho\pi})
           \frac{e^{-\left|\Delta t\right|/\tau}}{4\tau}} \\
        &&\hspace{1.3cm}\times\,\bigg[1+Q_{\rm tag}
             (\Srhopi \pm \dSrhopi)\sin(\Delta m_d \Delta t)\nonumber\\[-0.1cm]
        &&\hspace{1.3cm}\phantom{\times\,\bigg[1}
            -Q_{\rm tag}
                (\Crhopi \pm \dCrhopi)\cos(\Delta m_d \Delta t)\bigg]\;,\nonumber
\end{eqnarray}
where $Q_{\rm tag}= 1(-1)$ when the tagging meson $\Bz_{\rm tag}$
is a $\Bz(\Bzb)$, $\tau$ is the mean
\Bz lifetime, and $\Delta m_d$ the mixing frequency due to the
eigenstate mass difference. The observables $A_{\rm CP}^{\rho\pi}$, $\Srhopi$, $\dSrhopi$, $\Crhopi$ and~$\dCrhopi$ can be written in terms of the amplitudes defined in Eq.~(\ref{eq:b0rho+pi-}):
\begin{eqnarray}
A_{\rm CP}^{\rho\pi} = \frac{|\Apm|^2 + |\Apmb|^2 - |\Amp|^2 - |\Ampb|^2}
                  {|\Apm|^2 + |\Apmb|^2 + |\Amp|^2 + |\Ampb|^2} \,,\nonumber\\
\Crhopi = \frac{1 - |\lamCP|^2}{1 + |\lamCP|^2}\,, \hspace{1cm}
\dCrhopi = \frac{1 - |\lamFL|^2}{1 + |\lamFL|^2}\,,\\
\Srhopi = \frac{2 {\rm Im}\lamCP}{1 + |\lamCP|^2}\,, \hspace{1cm}
\dSrhopi = \frac{2 {\rm Im}\lamFL}{1 + |\lamFL|^2}\,.\hspace{0.75mm}\nonumber
\end{eqnarray}
The quantity~$A_{\rm CP}^{\rho\pi}$ measures time-integrated and $B$-flavour-integrated direct CP~violation. Flavour-dependent CP~violation is parameterised by~$\Crhopi$. $\Srhopi$~measures mixing-induced CP~violation related to the angle~$\alpha$. The quantities $\dCrhopi$~and~$\dSrhopi$ are insensitive to CP~violation. $\dCrhopi$~is a measure of the self-taggedness of the $\Bz/\Bzb \to \rho^{\pm} \pi^{\mp}$ decay, and $\dSrhopi$~is related to the strong phase differences between the different $T$~and~$P$ amplitudes in Eq.~(\ref{eq:b0rho+pi-}).

We adopt the definition of the {\it effective weak angles} $\alpha_{\rm eff}^{+-}$~and~$\alpha_{\rm eff}^{-+}$ given in Ref.~\cite{bib:Jerome}:
\begin{equation}
2\alpha_{\rm eff}^{+-} \equiv \arg{\kappa^{-+}}\,, \hspace{0.5cm} 2\alpha_{\rm eff}^{-+} \equiv \arg{\kappa^{+-}}\,.
\end{equation}
The relation between the observables in Eq.~(\ref{eq:thTime}) and these effective weak angles can be written as 
\begin{eqnarray}
\Srhopi + \dSrhopi = \sqrt{1-(\Crhopi + \dCrhopi)^2} \, \sin(2 \alpha_{\rm eff}^{+-}+\hat{\delta}) \,,\\
\Srhopi - \dSrhopi = \sqrt{1-(\Crhopi - \dCrhopi)^2} \, \sin(2 \alpha_{\rm eff}^{-+}-\hat{\delta}) \,,
\end{eqnarray}
where $\hat{\delta} = \arg(\Amp {\Apm}^*)$. The parameter~$\hat{\delta}$ is related to the phase between the amplitudes $A(\Bz\rar\rho^+\pi^-)$ and $A(\Bz\rar\rho^-\pi^+)$. It cannot be determined in a quasi-two-body analysis as one restricts oneself to regions of the Dalitz plot that are dominated by either $\rho^+ \pi^-$ or $\rho^- \pi^+$. Information from the region of the Dalitz plot where $\rho^+ \pi^-$ and $\rho^- \pi^+$ interfere is needed to determine~$\hat{\delta}$. In the case of negligible penguin amplitudes, the effective weak angles are equal to the angle~$\alpha$ of the UT.

\subsection{\babar\ measurement of CP-violating asymmetries in $B^0 \to \rho^{\pm} \pi^{\mp}$ and $B^0 \to \rho^- K^+$}
\label{sec:measurement}
A first preliminary measurement of the parameters $A_{\rm CP}^{\rho\pi}$, $\Srhopi$, $\dSrhopi$, $\Crhopi$ and~$\dCrhopi$ has been presented by the \babar\ Collaboration at ICHEP~2002~\cite{bib:RhoPiICHEP}. An improved version of the analysis with significantly reduced systematic uncertainties has recently been finalised~\cite{bib:RhoPiPRL}. This update also includes precise measurements of the branching fractions $\BRall={\cal B}(\Bz/\Bzb \rightarrow \rho^{\pm} \pi^{\mp})$ and $\BRrhoK={\cal B}(\Bz/\Bzb \rightarrow \rho^{\pm} K^{\mp})$.

\babar's internally reflecting ring-imaging Cherenkov detector provides good separation between kaons and pions from $B \rightarrow \rho \pi$ and $B \rightarrow \rho K$ (the typical separation depends on the particle's momentum and varies from $8 \sigma$ at~$2 \; {\rm GeV}/c$ to $2.5 \sigma$ at $4 \; {\rm GeV}/c$). To treat the small overlap at higher momenta in an optimal way, a simultaneous analysis of the $\rho \pi$~and~$\rho K$ states is performed. The time-dependence of the $B \rightarrow \rho K$ decays is given by Eq.~(\ref{eq:thTime}) with $\CrhoK=\SrhoK=\dSrhoK = 0$ and $\dCrhoK = -1$ as the $B \rightarrow \rho K$ decay is self-tagging. As the $\rho \pi / \rho K$~branching fractions are small~$[{\cal O}(10^{-5})]$ and the $\rho$~resonance is broad, the $\rho \pi$~and~$\rho K$ channels suffer from large backgrounds. Continuum $e^+ e^- \rightarrow q \overline{q} \; (q=u,d,s,c)$ events are the dominant source of background. As these events are very abundant, their properties can be extracted from the data simultaneously with the parameters of the $\rho \pi / \rho K$~signals. Cross-feed from other $B$~decays (both charmless and $b \rightarrow c$~decays) is potentially more dangerous, as some of the corresponding final states are very similar to the signal, and these backgrounds are hidden, together with the signal, in a much larger number of continuum background events. $B$~candidates are identified using the beam-energy substituted mass~$m_{\rm ES}$~\cite{bib:RhoPiICHEP} and the difference~$\Delta E$ between the centre-of-mass (CM) energy of the $B$~candidate and~$\frac{\sqrt{s}}{2}$, where $\sqrt{s}$~is the total CM~energy. $\Delta E$~also provides good discrimination between signal and background from higher-multiplicity $B$~decays, whose branching fraction is not well known in many cases. To enhance the discrimination between signal and continuum, a neural network~(NN) is used to combine four discriminating variables: two event-shape variables (continuum events tend to be jet-like, while $B$~events are more spherical), the mass of the $\rho$~candidate, and~$\cos{\theta_{\pi}}$, where $\theta_{\pi}$~is defined as the angle between the $\pi^0$~momentum and the negative $B$~momentum in the $\rho$~rest frame. The $\rho\pi$~and~$\rho K$ event yields as well as the observables discussed above are determined in a simultaneous likelihood fit. The event variables used as input to the fit are $m_{\rm ES}$, $\Delta E$, NN output, $\pi/K$~Cherenkov angle and $\Delta t$ with its uncertainty~$\sigma_{\Delta t}$. The likelihood function contains contributions for signal, continuum and $B$-related backgrounds. 24~parameters that provide an empirical description of the continuum events are also free to vary in the fit. Cross-feed from other $B$~decays  is studied using simulated events. The branching fractions of unmeasured decay channels are estimated within conservative error ranges. More than 100~charmless decay modes have been studied, and the 31~which contribute to the final event sample, plus the contribution from $b \rightarrow c$ decays, are taken into account in the likelihood function. Backgrounds from two-, three-, and four-body decays to $\rho \pi$ are dominated by $B^+\rightarrow\pi^+\pi^0$, $B^+\rightarrow\rho^0\pi^+$, and longitudinally polarized $B^0\rightarrow\rho^+\rho^-$ decays, respectively. The $\rho K$ sample receives dominant two-body background from $B^+\rightarrow K^+\pi^0$ and three- and four-body background from $B\rightarrow K{^\ast}\pi$ and higher kaonic resonances, estimated from inclusive $B\rightarrow K\pi\pi$ measurements.
Using $89 \times 10^6$ $B\overline{B}$~pairs collected at the~$\Upsilon(4S)$, the \babar\ Collaboration finds $428 \pm 34$ ($120 \pm 21$) $\rho^{\pm} \pi^{\mp}$ ($\rho^{\pm} K^{\mp}$)~events and measures~\cite{bib:RhoPiPRL}:
\begin{eqnarray*}
\begin{array}{rcl}
\BRall &=&
        (22.6 \pm 1.8
        \pm2.2)\times 10^{-6}\,, \\
\BRrhoK &=&
        (7.3^{\,+1.3}_{\,-1.2}
        \pm1.3)\times 10^{-6}\,, \\
\end{array}\\
A_{\rm CP}^{\rho \pi} =  - 0.18\pm 0.08 \pm{0.03}\,,  \,\,A_{\rm CP}^{\rho K}  = 0.28 \pm 0.17 \pm 0.08\,,\\
\Crhopi = \phantom{-}0.36 \pm 0.18 \pm 0.04\,, \,\,\,\Srhopi =  0.19\pm 0.24\pm 0.03\,,\\
\dCrhopi = \phantom{-}0.28^{+0.18}_{-0.19}\pm 0.04\,,  \,\,\dSrhopi = 0.15\pm 0.25\pm 0.03\,,
\end{eqnarray*}
where the first errors are statistical and the second systematic. The dominant systematic uncertainties are due to the uncertainties in the model of the $B$-related backgrounds, and possible interference between the doubly-Cabibbo-suppressed $\overline{b} \rightarrow \overline{u} c \overline{d}$ amplitude with the Cabibbo-favoured $b \rightarrow c \overline{u} d$ amplitude for tag-side $B$~decays~\cite{bib:DCSD}. Other important uncertainties, especially in the branching fraction measurements, are due to the uncertainties on the shapes of the reference distributions for signal used in the likelihood function, and to fit biases caused by small effects that are not included in the likelihood model. While the branching fraction measurements are limited by systematic uncertainties, the other measurements will remain statistics limited for quite some time. The uncertainties due to $B$-related backgrounds are expected to improve as more experimental constraints on the corresponding decay modes become available. These results are consistent with CP~conservation at the $2.3 \sigma$~level.

\subsection{Expected constraints on~$\alpha$}
\label{sec:constraints}
\begin{figure*}
\hbox to\hsize{\hss
\includegraphics[width=0.33\hsize]{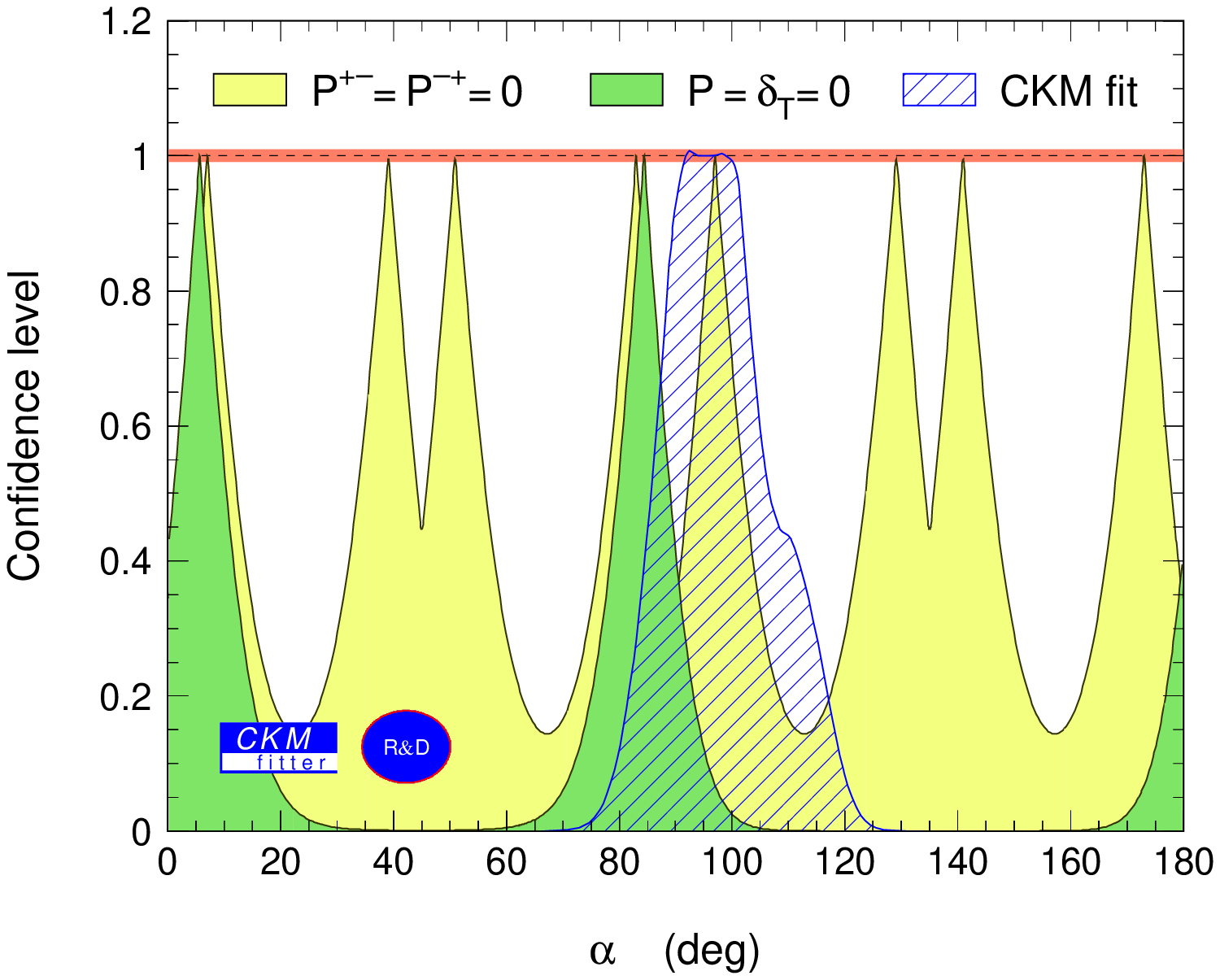}
\hss
\includegraphics[width=0.33\hsize]{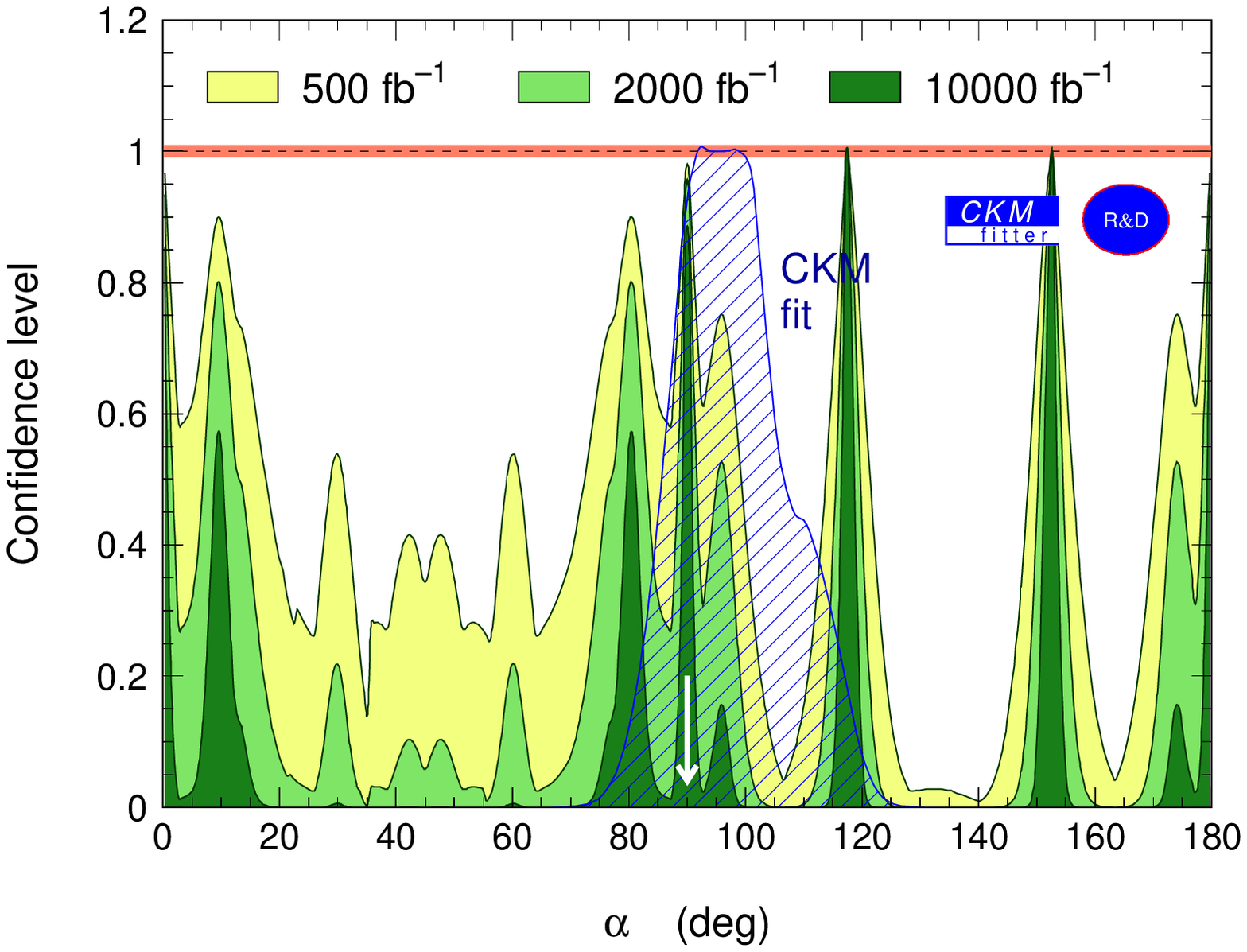}
\hss
\includegraphics[width=0.33\hsize]{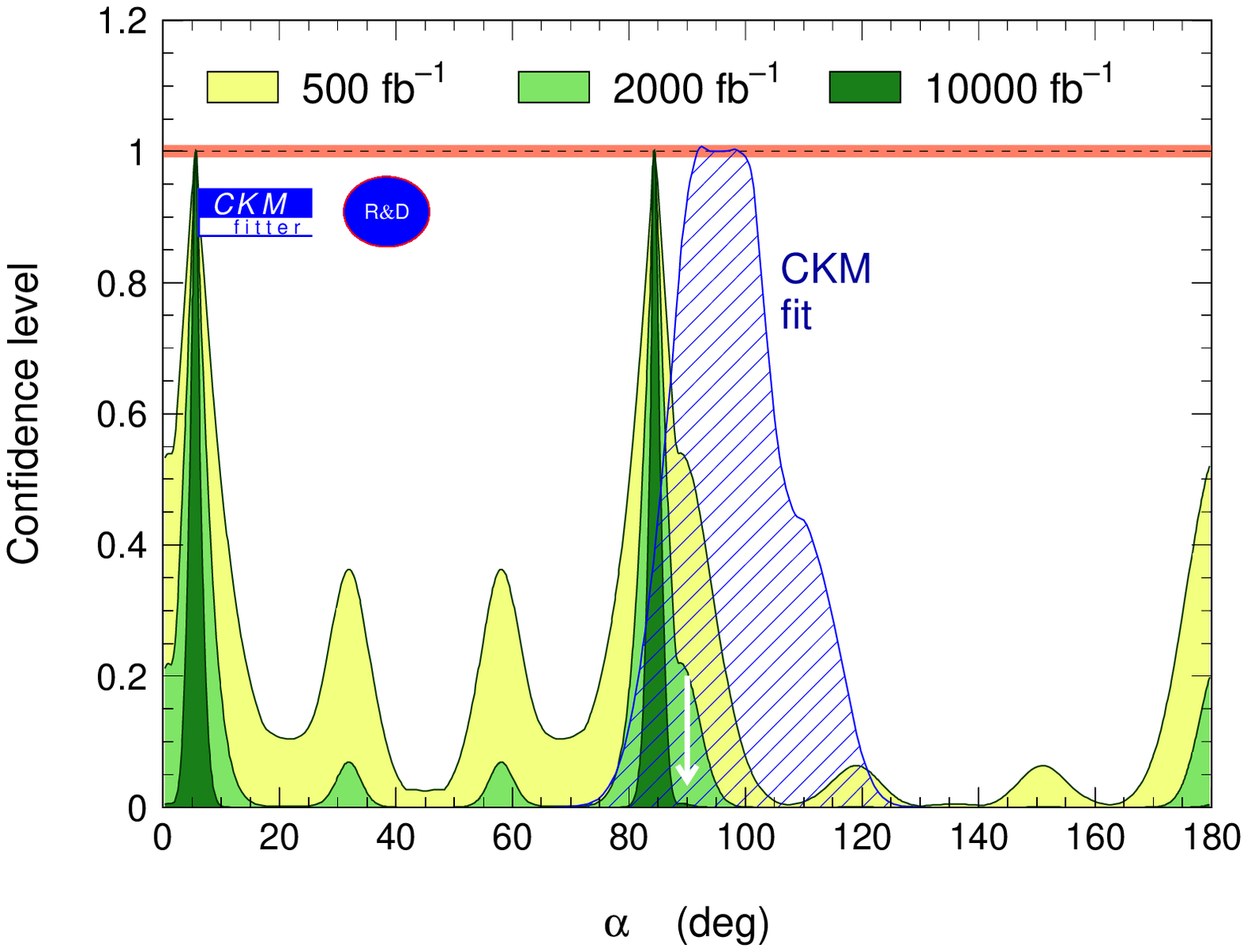}
\hss
}
\vskip-4.5cm
(a)~~~~~~~~~~~~~~~~~~~~~~~~~~~~~~~~~~~~~~~~~~~~~~~~~~~~~~~~~~~~(b)~~~~~~~~~~~~~~~~~~~~~~~~~~~~~~~~~~~~~~~~~~~~~~~~~~~~~~~~~~~~~(c)
\vskip4cm
\caption{Expected constraints on~$\alpha$ using SU(2) flavour symmetry for different scenarios (see text).}
\label{fig:constraints}
\end{figure*}

Figure~\ref{fig:constraints}(a) shows the constraints on~$\alpha$ obtained using the \babar\ measurement and the assumption that penguin amplitudes are negligible ($\Ppm=\Pmp=0$). The eightfold ambiguity is due to the presence of the unknown strong phase~$\hat{\delta}$. This figure also shows the constraints obtained when, in addition, the strong phase~$\hat{\delta}$ is assumed to be zero ($P=\delta_{\rm T}=0$). This exercise is purely academic as the penguin contributions are not expected to be negligible and the strong phase~$\hat{\delta} = \arg(\Tmp {\Tpm}^*)$ is unknown. It is shown here to illustrate the accuracy that could be obtained if the penguin amplitudes and the strong phases of the tree diagrams were known (e.g. from theory) and to set the scale for the discussion that follows. In this case, $\alpha$~could be determined up to a twofold ambiguity with an accuracy of $7^\circ$~per solution. This is better than the indirect constraints on~$\alpha$ from the standard global fit of the CKM~matrix (see Fig.~\ref{fig:constraints}(a))~\cite{bib:CKMfit}.

With the assumption of SU(2) flavour symmetry, the penguins can be constrained using experimental information on the other $B \rightarrow \rho\pi$ charge combinations. The key observation is that gluonic $b \rightarrow d$~penguins are pure~$\Delta I=\frac{1}{2}$, while tree-level $b \rightarrow u\overline{u}d$ decays have $\Delta I=\frac{3}{2}$~and~$\Delta I=\frac{1}{2}$ components. With the definitions $\Apz = A(B^+ \rightarrow \rho^+ \pi^0)$, $\Azp = A(B^+ \rightarrow \rho^0 \pi^+)$ and $\Azz = A(B^0 \rightarrow \rho^0 \pi^0)$ this leads to the two pentagon relations
\begin{eqnarray}
\label{eq:pentagon}
\sqrt{2}\left(\Apz+\Azp\right) &=& 2\Azz+\Apm +\Amp~, \\
\sqrt{2}\left(\Apzb+\Azpb\right) &=& 2\Azzb+\Apmb +\Ampb~;
\end{eqnarray}
see~\cite{bib:PhysBook} for the details of the isospin decomposition. The effects of electroweak penguins and electromagnetic transitions, which can have $\Delta I=\frac{3}{2}$~components, are neglected here. This leads to 12~unknowns (6~complex amplitudes $\Tpm$, $\Ppm$, $\Tmp$, $\Pmp$, $\Tzz$ and $\Tpz$, and $\alpha$, minus one arbitrary global phase) and 13~observables (see Table~\ref{tab:observables}).
\begin{table*}
\begin{center}
\begin{tabular}{llll}
Mode                                   & Observables                                        & $N_{\rm observables}$                & Experimental status\\
\hline
$\Bz \rightarrow \rho^{\pm} \pi^{\mp}$ & $\BRall$, $\Crhopi$, $\dCrhopi$, $\Srhopi$, $\dSrhopi$, $A_{\rm CP}^{\rho\pi}$ & 6 & Measured~\cite{bib:RhoPiICHEP,bib:RhoPiPRL}.\\
$B^+ \rightarrow \rho^+ \pi^0$         & $\BRpz$, $A_{\rm CP}^{+0}$                         & 2 & Not observed. Limit on $\BRpz$:~\cite{bib:rhopibfCLEO}.\\
$B^+ \rightarrow \rho^0 \pi^+$         & $\BRzp$, $A_{\rm CP}^{0+}$                         & 2 & Only $\BRzp$ measured~\cite{bib:rhopibfCLEO,bib:rhopibfBelle,bib:Osaka}.\\
$\Bz \rightarrow \rho^0 \pi^0$         & $\BRzz$, $\Czz$, $\Szz$                            & 3 & Not observed. Limits on $\BRzz$:~\cite{bib:rhopibfCLEO,bib:rhopibfBelle}.\\
\end{tabular}
\end{center}
\caption{Observable-counting in the SU(2) analysis.}
\label{tab:observables}
\end{table*}
The isospin analysis is not yet feasible with present statics from the $B$-factories (some of the relevant decay modes have not even been observed, Table~\ref{tab:observables}). A rough extrapolation into the future is presented in~\cite{bib:LALnote}. With an integrated luminosity of 500~fb$^{-1}$ the experimental situation might look like in Table~\ref{tab:outlook}. One of the infinite number of sets of values for the 12~unknowns that reproduce the \babar\ measurement has been chosen to obtain the central values. The uncertainties are educated guesses based on the uncertainties of present measurements and searches, (optimistically) assumed to scale with~$1/\sqrt{N}$. The constraint on~$\alpha$ obtained in this scenario is shown in Fig.~\ref{fig:constraints}(b). This figure also shows the constraint for even higher luminosities of~2~ab$^{-1}$ (achievable by the first generation $B$-factories ?) and~10~ab$^{-1}$ (Super~$B$-factory). Very high luminosities are needed to obtain meaningful constraints on~$\alpha$, and even at high luminosities several solutions exist. Figure~\ref{fig:constraints}(c) shows a slightly different scenario where ${\cal B}(B^0 \rightarrow \rho^0 \pi^0)$ is assumed to be significantly smaller than in the first scenario (below the experimental sensitivity). In this case, the SU(2) analysis provides meaningful constraints above~2~ab$^{-1}$.
\begin{table*}
\begin{center}
$$
\begin{array}{rclrclrcl}
        \BRall  &=&     (22.6 \pm 1.0)\;10^{-6}~, &
        A_{\rm CP}^{\rho\pi}    &=&     -0.18 \pm 0.04~, &
        \Czz    &=&     \phantom{-}0.56\pm 0.35~,\\[0.1cm]
        \BRzz   &=&     (\phantom{2}0.9  \pm 0.3)\;10^{-6}~, &
        \Crhopi &=&     \phantom{-}0.36 \pm 0.08~, &
        \Szz    &=&     \phantom{-}0.82 \pm 0.50~,\\[0.1cm]
        \BRpz   &=&     (\phantom{2}8.1  \pm 1.0)\;10^{-6}~, &
        \dCrhopi&=&     \phantom{-}0.28 \pm 0.08~, &
        A_{\rm CP}^{+0}  &=&     \phantom{-}0.02\pm 0.07~, \\[0.1cm]
        \BRzp   &=&     (8.7 \pm 0.5)\;10^{-6}~, &
        \Srhopi &=&     \phantom{-}0.19 \pm 0.10~, &
        A_{\rm CP}^{0+}  &=&     \phantom{-}0.36 \pm 0.05~, \\[0.1cm]
        &&&
        \dSrhopi&=&     \phantom{-}0.15 \pm 0.11~,
\end{array}
$$
\end{center}
\caption{Extrapolation to an integrated luminosity of 500~fb$^{-1}$.}
\label{tab:outlook}
\end{table*}

\section{Time-dependent Dalitz plot analysis}
\label{sec:Dalitz}
This analysis allows the measurement of~$\alpha$ without ambiguities even in the presence of non-negligible penguin contributions. The $B^0 \rightarrow \pi^+ \pi^- \pi^0$ amplitude is written as sum over the resonant intermediate states~$\rho^+ \pi^-$, $\rho^- \pi^+$ and~$\rho^0 \pi^0$:
\begin{equation}
\label{eq:DalitzAmplitude}
A = f^{+-}\Apm + f^{-+}\Amp + f^{00}\Azz~.
\end{equation}
The {\it form factors} $f^{+-}$, $f^{-+}$, and~$f^{00}$ describe the full kinematics of the three-pion final state. They include the dominant $\rho(770)$ as well as higher-mass states. An estimate of the form factors can be obtained from $e^+ e^- \rightarrow \pi^+ \pi^-$ data and from $\tau^+ \rightarrow \pi^+ \pi^0 \nu_{\tau}$ decays. Toy Monte Carlo studies with samples of 500~events of pure signal (this corresponds to approximately 100~fb$^{-1}$) with realistic $\Delta t$~resolution and flavour tagging give typical statistical uncertainties on~$\alpha$ of the order of $6^\circ - 10^\circ$~\cite{bib:Sandrine}, depending on the values of the different amplitudes in Eq.~(\ref{eq:DalitzAmplitude}), and for values consistent with the quasi-two-body result from \babar. In these toy fits, all (complex) amplitudes ($\Tpm$, $\Tmp$, $\Ppm$, $\Pmp$ and $\Tzz$) were free to vary and determined from the toy data. The inclusion of a realistic number of continuum background events increases the statistical error by about~$25 \; \%$. As in the quasi-two-body analysis, the impact of $B$-related backgrounds is important. Some of these backgrounds tend to accumulate in the interference regions of the Dalitz plot which are essential for the Dalitz plot analysis. The inclusion of realistic $B$-related backgrounds in the toys increases the statistical error by about~$30 \; \%$. Shifts of up to~$40^\circ$ in the fitted value of~$\alpha$ are observed in toy experiments where realistic $B$-backgrounds are simulated but not taken into account in the fit~\cite{bib:Sandrine}. While $B$-backgrounds will be modelled in any real Dalitz plot analysis, the systematic uncertainties due to imperfections in the $B$-background model are potentially large. Various other issues will need to be addressed. These include the precise shape of the form factors (moduli and phases). The impact of possible contributions to $B^0 \rightarrow \pi^+ \pi^- \pi^0$ from very broad resonances and of non-resonant contributions needs to be evaluated. It could be possible to include a corresponding term in the amplitude~(\ref{eq:DalitzAmplitude}) and fit for its normalisation in the data. A more detailed discussion of these issues can be found, e.g., in~\cite{bib:Sophie,bib:Polosa,bib:Oller,bib:GM,bib:TG}. This list of issues is not intended to be exhaustive.

\section{Conclusion}
The first time-dependent quasi-two-body analyses of $B \rightarrow \rho \pi$ are becoming available from the $B$-factories. With SU(2) flavour symmetry, these measurements can be used to constrain the angle~$\alpha$. These constraints on~$\alpha$ are only expected to be significant if the branching fraction of the yet unobserved decay $B^0 \rightarrow \rho^0 \pi^0$ turns out to be well below the current experimental sensitivity. The quasi-two-body analyses are also an important first step towards a time-dependent Dalitz plot analysis. This analysis could provide strong constraints on~$\alpha$ with the luminosities expected from the $B$-factories, but various experimental and theoretical issues still need to be addressed.

\section*{Acknowledgments}
It is a pleasure to thank my (former) \babar\ collaborators for several years of exciting and fruitful collaboration, and our PEP-II colleagues for the excellent luminosity. I am indebted to the authors of H\"ocker {\it et al.}~\cite{bib:LALnote} for valuable discussions and comments.

\end{document}